\documentclass[twocolumn]{emulateapj}

\begin{document}

\title{Discovery of Main-Belt Comet P/2006 VW$_{139}$ by Pan-STARRS1}

\author{
%Contributors
Henry H.\ Hsieh\altaffilmark{1,a},
Bin Yang\altaffilmark{1},~%UH
Nader Haghighipour\altaffilmark{1},~%UH
Heather M.\ Kaluna\altaffilmark{1},~%UH
Alan Fitzsimmons\altaffilmark{2},~%QUB
Larry Denneau\altaffilmark{1},~%UH
Bojan Novakovi\'c\altaffilmark{3},~%Belgrade
%Observers
Robert Jedicke\altaffilmark{1},~%UH
Richard J.\ Wainscoat\altaffilmark{1},~%UH
James D.\ Armstrong\altaffilmark{1,4},~%UH
Samuel R.\ Duddy\altaffilmark{5},~%Kent
Stephen C.\ Lowry\altaffilmark{5},~%Kent
Chadwick A.\ Trujillo\altaffilmark{6},~%Gemini
Marco Micheli\altaffilmark{1},~%UH
Jacqueline V.\ Keane\altaffilmark{1},~%UH
Laurie Urban\altaffilmark{1},~%UH
Timm Riesen\altaffilmark{1},~%UH
Karen J.\ Meech\altaffilmark{1},~%UH
Shinsuke Abe\altaffilmark{7},~%NCU
Yu-Chi Cheng\altaffilmark{7},~%NCU
%ISS
Wen-Ping Chen\altaffilmark{7},~%NCU
Mikael Granvik\altaffilmark{8},~%Hels
Tommy Grav\altaffilmark{9},~%PSI
Wing-Huen Ip\altaffilmark{7},~%NCU
Daisuke Kinoshita\altaffilmark{7},~%NCU
Jan Kleyna\altaffilmark{1},~%UH
Pedro Lacerda\altaffilmark{2,b},~%QUB
Tim Lister\altaffilmark{4},~%LCOGT
Andrea Milani\altaffilmark{10},~%Pisa
David J.\ Tholen\altaffilmark{1},~%UH
Peter Vere{\v s}\altaffilmark{1},~%UH
Carey M.\ Lisse\altaffilmark{11},~%APL
Michael S.\ Kelley\altaffilmark{12},~%UMD
Yanga R.\ Fern\'andez\altaffilmark{13},~%UCF
Bhuwan C.\ Bhatt\altaffilmark{14},~%India
Devendra K.\ Sahu\altaffilmark{14},~%India
%Builders
Nick Kaiser\altaffilmark{1},~%UH
K.~C.\ Chambers\altaffilmark{1},~%UH
Klaus W.\ Hodapp\altaffilmark{1},~%UH
Eugene A.\ Magnier\altaffilmark{1},~%UH
Paul A.\ Price\altaffilmark{15},~%Princeton
John L.\ Tonry\altaffilmark{1}~%UH
%W. S. Burgett\altaffilmark{1}
%J. N. Heasley\altaffilmark{1},
%R.-P. Kudritzki\altaffilmark{1},
%G. A. Luppino\altaffilmark{1},
%R. H. Lupton\altaffilmark{3},
%D. G. Monet\altaffilmark{4},
%J. S. Morgan\altaffilmark{1},
%P. M. Onaka\altaffilmark{1},
}
%\affil{
\altaffiltext{1}{Institute for Astronomy, University of Hawaii, 2680 Woodlawn Drive, Honolulu, HI 96822, USA}
\altaffiltext{2}{Astrophysics Research Centre, Queens University Belfast, Belfast BT7 1NN, United Kingdom}
\altaffiltext{3}{Department of Astronomy, Faculty of Mathematics, University of Belgrade, Studentski trg 16, 11000 Belgrade, Serbia}
\altaffiltext{4}{Las Cumbres Observatory Global Telescope Network, Inc., 6740 Cortona Dr.\ Suite 102, Santa Barbara, CA 93117 USA}
\altaffiltext{5}{Centre for Astrophysics and Planetary Science, The University of Kent, Canterbury CT2 7NH, United Kingdom}
\altaffiltext{6}{Gemini Observatory, Northern Operations Center, 670 N.\ AÔohoku Place, Hilo, HI 96720, USA}
\altaffiltext{7}{Institute of Astronomy, National Central University, 300 Jhongda Rd, Jhongli 32001, Taiwan}
\altaffiltext{8}{Department of Physics, P.O. Box 64, 00014 University of Helsinki, Finland}
\altaffiltext{9}{Planetary Science Institute, 1700 East Fort Lowell, Suite 106, Tucson, AZ 85719}
\altaffiltext{10}{Dipartimento di Matematica, Universit{\`a} di Pisa, Largo Pontecorvo 5, 56127 Pisa, Italy}
\altaffiltext{11}{Planetary Exploration Group, Space Department, Johns Hopkins University Applied Physics Laboratory, Laurel, MD 20723, USA}
\altaffiltext{12}{Department of Astronomy, University of Maryland, College Park, MD 20742, USA}
\altaffiltext{13}{Department of Physics, University of Central Florida, 4000 Central Florida Blvd., Orlando, FL 32816, USA}
\altaffiltext{14}{Indian Institute of Astrophysics, CREST Campus, Block-II, Koramangala, Sarjapur Road, Bangalore 560034, India}
\altaffiltext{15}{Department of Astrophysical Sciences, Peyton Hall, Princeton University, Princeton, 08544, USA}
\altaffiltext{a}{Hubble Fellow}
\altaffiltext{a}{Michael West Fellow}
%}
\email{hsieh@ifa.hawaii.edu}

\slugcomment{ApJ Letters - Submitted, 2012-01-20; Accepted, 2012-02-09}

\begin{abstract}
Main belt asteroid (300163) 2006~VW$_{139}$ (later designated P/2006~VW$_{139}$) was discovered to exhibit comet-like activity by the Pan-STARRS1 survey telescope using automated point-spread-function analyses performed by PS1's Moving Object Processing System.  Deep follow-up observations show both a short ($\sim10''$) antisolar dust tail and a longer ($\sim60''$) dust trail aligned with the object's orbit plane, similar to the morphology observed for another main-belt comet, P/2010 R2 (La Sagra), and other well-established comets, implying the action of a long-lived, sublimation-driven emission event.  Photometry showing the brightness of the near-nucleus coma remaining constant over $\sim30$~days provides further evidence for this object's cometary nature, suggesting it is in fact a main-belt comet, and not a disrupted asteroid.  A spectroscopic search for CN emission was unsuccessful, though we find an upper limit CN production rate of $Q_{\rm CN}<1.3\times10^{24}$~mol~s$^{-1}$, from which we infer a water production rate of $Q_{\rm H_2O}<10^{26}$~mol~s$^{-1}$.  We also find an approximately linear optical spectral slope of 7.2\%/1000\AA , similar to other cometary dust comae.  Numerical simulations indicate that P/2006~VW$_{139}$ is dynamically stable for $>100$~Myr, while a search for a potential asteroid family around the object reveals a cluster of 24 asteroids within a cutoff distance of 68~m~s$^{-1}$.  At 70~m~s$^{-1}$, this cluster merges with the Themis family, suggesting that it could be similar to the Beagle family to which another main-belt comet, 133P/Elst-Pizarro, belongs.
%Though we do not detect CN emission, the morphological and photometric evidence strongly suggest that the observed activity is cometary in nature, and thus we conclude that P/2006~VW$_{139}$ is likely a main-belt comet and not a disrupted asteroid.
\end{abstract}


\newpage

\section{INTRODUCTION}

Main-belt comets (MBCs) exhibit cometary activity indicative of ice sublimation yet are dynamically indistinguishable from main-belt asteroids \citep{hsi06}.  Much of the current interest in studying MBCs lies in the possible role of main-belt objects in the primordial delivery of terrestrial water \citep[e.g.,][]{mor00}.  Of the first five known MBCs, one (176P/LINEAR) was discovered via a targeted search \citep{hsi09}, while the other four were discovered serendipitously or by untargeted surveys \citep{els96,rea05,gar08,nom10}.  Using MBCs as tracers of ice in the asteroid belt to ascertain its potential for water delivery will require a much larger sample of these objects, and as such, discovering more MBCs, ideally in untargeted surveys \citep[e.g.,][]{gil09,son11}, is a high priority.

Recently, main-belt objects P/2010 A2 (LINEAR) and (596) Scheila have exhibited comet-like dust emission, but these events are likely due to impact-generated ejecta clouds \citep{jew10,jew11,sno10,bod11,yan11,ish11b,hai12}.  As such, these objects are better characterized as disrupted asteroids.  \citet{hsi12a} considered the problem of distinguishing MBCs and disrupted asteroids, and concluded that recurrent activity separated by periods of inactivity to be the strongest observable indicator that activity is sublimation-driven, particularly when the timing of active episodes corresponds closely to an object's orbital period \citep[cf.\ 133P/Elst-Pizarro and 238P/Read;][]{hsi04,hsi10,hsi11b}.  Evaluating this criterion requires monitoring an object for many years, however.  When observations of only one active episode for an object are available, \citet{hsi12a} suggested that steady or increasing activity and morphological indicators could be used as preliminary evidence of sublimation.  Direct spectroscopic detection of sublimation products (i.e., gas) in an MBC would immediately confirm its cometary nature. Given the difficulty of obtaining sufficiently high-quality spectroscopy of such distant and weakly-active comets and of timing observations to coincide with peak gas production though, the absence of gas detections to date \citep{jew09,hsi12b} should not be considered confirmation of the absence of sublimation.

\section{OBSERVATIONS\label{observations}}

Observations of activity in P/2006 VW$_{139}$ were first obtained by the 1.8~m Pan-STARRS1 (PS1) wide-field synoptic survey telescope on Haleakala.  PS1 employs a $3.2\degr\times3.2\degr$ 1.4 gigapixel camera, consisting of a mosaic of 60 orthogonal transfer arrays, each comprising 64 $590\times598$~pixel CCDs, and Sloan Digital Sky Survey (SDSS) $i'$- and $z'$-band-like filters designated $i_{\rm P1}$ and $z_{\rm P1}$ \citep{ton12}.

Follow-up imaging was performed using the 2.0~m Faulkes Telescope North (FTN) on Haleakala and Faulkes Telescope South (FTS) at Siding Spring, the University of Hawaii (UH) 2.2~m telescope on Mauna Kea, the 1.8~m Perkins Telescope (PT) at Lowell Observatory, the 2.0~m Himalayan Chandra Telescope (HCT) at the Indian Astronomical Observatory on Mt.\ Saraswati, the 4.2~m William Herschel Telescope (WHT; Program SW2011b20) at La Palma, the 3.54~m New Technology Telescope (NTT; Program 185.C-1033(K)) operated by the European Southern Observatory (ESO) at La Silla, and the Lulin One-meter Telescope (LOT) in Taiwan.
We employed $4096\times4096$~pixel Fairchild CCDs and either SDSS or Bessell filters for Faulkes observations, 
a $2048\times2048$~pixel Textronix CCD and Kron-Cousins filters for UH observations,
the Perkins ReImaging System (PRISM) and Kron-Cousins filters for Perkins observations,
a $2048\times2048$~pixel E2V CCD and Bessell filters for HCT observations,
the WHT's auxiliary-port camera \citep[ACAM;][]{ben08} and SDSS filters on the WHT, 
the ESO Faint Object Spectrograph and Camera \citep[EFOSC2;][]{buz84} and Bessell filters on the NTT, 
and a VersArray:1300B CCD \citep{kin05} and Bessell-like filters on the LOT.

We performed standard bias subtraction and flat-field reduction (using dithered twilight sky images) for all data, except those from PS1, using Image Reduction and Analysis Facility (IRAF) software.  PS1 data were reduced using the system's Image Processing Pipeline \citep[IPP;][]{mag06}.  Photometry of \citet{lan92} standard stars and field stars was performed by measuring net fluxes within circular apertures, with background sampled from surrounding circular annuli.  For data obtained under non-photometric conditions, absolute calibration was accomplished using SDSS field star magnitudes \citep{aih11}.  Conversion of $r'$-band PS1, FTN, FTS, and WHT photometry to $R$-band was accomplished using transformations derived by \citet{ton12} and by R.\ Lupton ({\tt http://www.sdss.org/}).  Comet photometry was performed using circular apertures with varying radii depending on the nightly seeing, where background statistics were measured in nearby, but non-adjacent, regions of blank sky to avoid dust contamination from the comet.
%At least five field stars in the comet images were also measured to correct for extinction variation during each night.

We also obtained longslit spectroscopy with ACAM (3500~${\rm \AA}$ to 9400~${\rm \AA}$) on the WHT, as well as with the Gemini Multi-Object Spectrographs \citep[GMOS; 3600~${\rm \AA}$ to 9400~${\rm \AA}$;][]{hoo04} on the 8-m Gemini North (GN; Program GN-2011B-Q-16) and Gemini South (GS; Program GS-2011B-Q-51) observatories.
For ACAM observations, we employed a $2\farcs0$-wide slit, giving a spectral resolution of $R\sim300$, aligned with the object's dust tail.
For GMOS observations, we employed B600 dispersers, 2$\times$2 binning, and $1\farcs0$-wide slits, giving $R\sim850$, aligned with the dust trail.  Reduction of GMOS data was performed using an IRAF package provided by Gemini.
%All spectroscopy was performed with slits aligned with the object's dust tail.

\section{RESULTS\label{results}}

\subsection{Discovery of Cometary Activity\label{discovery}}
\begin{figure}
\plotone{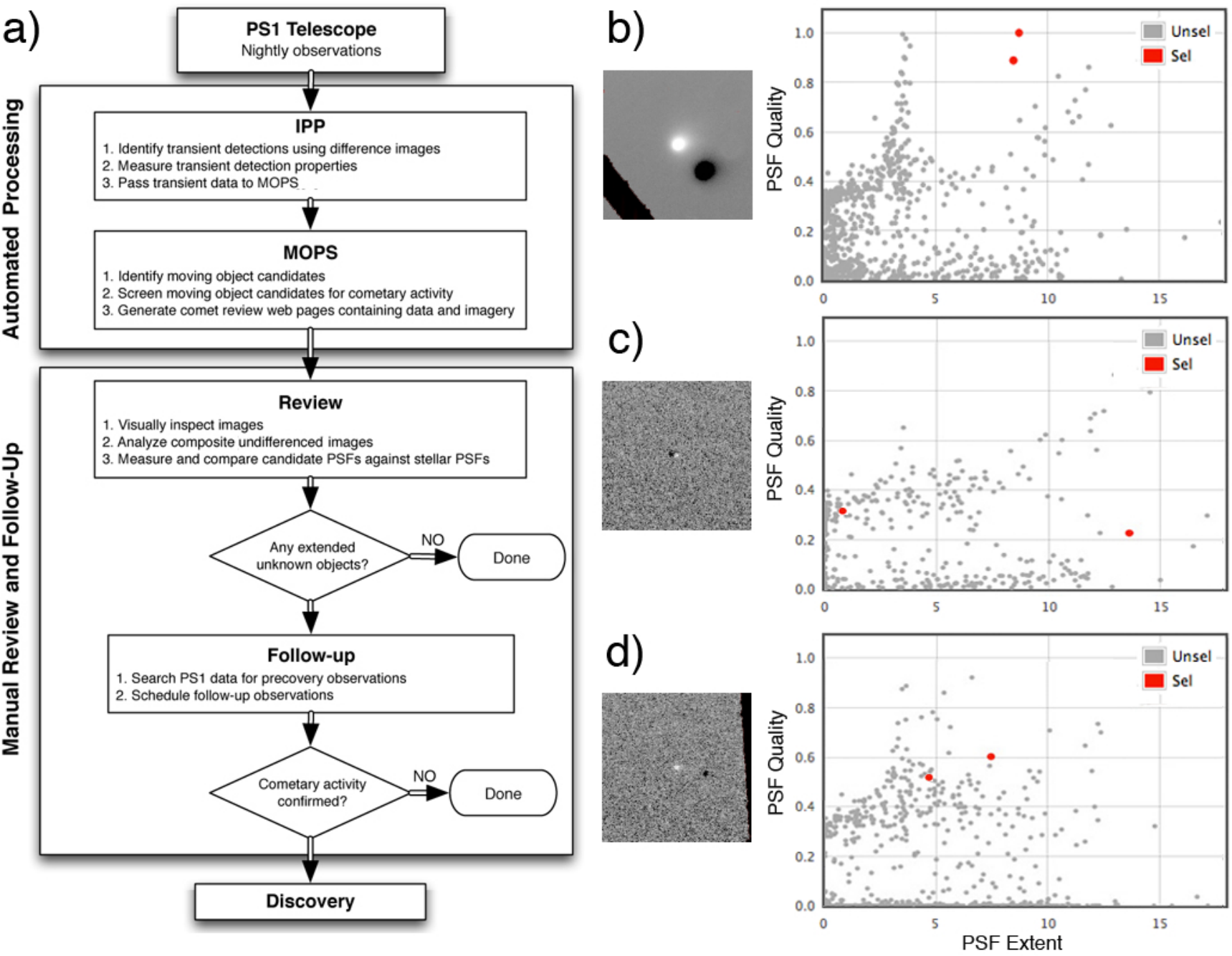}
\caption{\small 
(a) Flow chart detailing PS1 comet screening procedures.
(b) Difference image (left) and screening plot (right) for sample PS1 observations of 48P/Johnson where comet candidates (``selected'') are plotted as red dots and ``unselected'' detections (generally consisting of inactive asteroids and false detections) are plotted as gray dots.
(c) Image and screening plot for precovery observations of P/2006 VW$_{139}$ on 2011 August 30.
(d) Image and screening plot for discovery observations of P/2006 VW$_{139}$ on 2011 November 5.
  }
\label{fig_flowchart}
\end{figure}

All PS1 moving object detections are screened for potential cometary activity by automated point-spread function (PSF) analyses executed nightly by PS1's Moving Object Processing System (MOPS).  This screening process divides the measured second PSF moment of each transient source by the expected PSF width as determined from the median of all stellar PSFs in the field. This ``{\tt psfextent}'' parameter is then plotted as a function of a detection quality parameter (``{\tt psfquality}'') ranging from 0 to 1, roughly corresponding to a normalized signal-to-noise ratio.  Sources that show comet-like {\tt psfextent} parameters and {\tt psfquality} $>0.5$ (to screen out faint sources for which measured PSF moments are unreliable) are flagged for human inspection and possible observational follow-up (Figure~\ref{fig_flowchart}).

PS1 observations on 2011 November 5 showed that the PSF of the known main-belt asteroid, (300163) 2006~VW$_{139}$ (semimajor axis, $a=3.052$~AU; eccentricity, $e=0.201$; inclination, $i=3.24\degr$), had a FWHM of $1\farcs3$, while nearby stars had PSFs with FWHMs of $1\farcs0$ \citep{hsi11c}.  Precovery observations also obtained by PS1 showed that the object exhibited a similar PSF excess on 2011 August 30.  At the time, however, the object was too faint ({\tt psfquality} $<0.5$) and was not flagged for inspection.  In both the relatively shallow August and November PS1 data, no extended cometary features were seen.  Deeper follow-up observations (Table~\ref{obslog}), however, revealed the presence of both an antisolar dust tail and an orbit-aligned dust trail (Figure~\ref{fig_images_orbplot}; Section~\ref{morph_phot_results}), leading to the object's re-designation as P/2006~VW$_{139}$.

\subsection{Morphological \& Photometric Analysis\label{morph_phot_results}}

\begin{figure}
\plotone{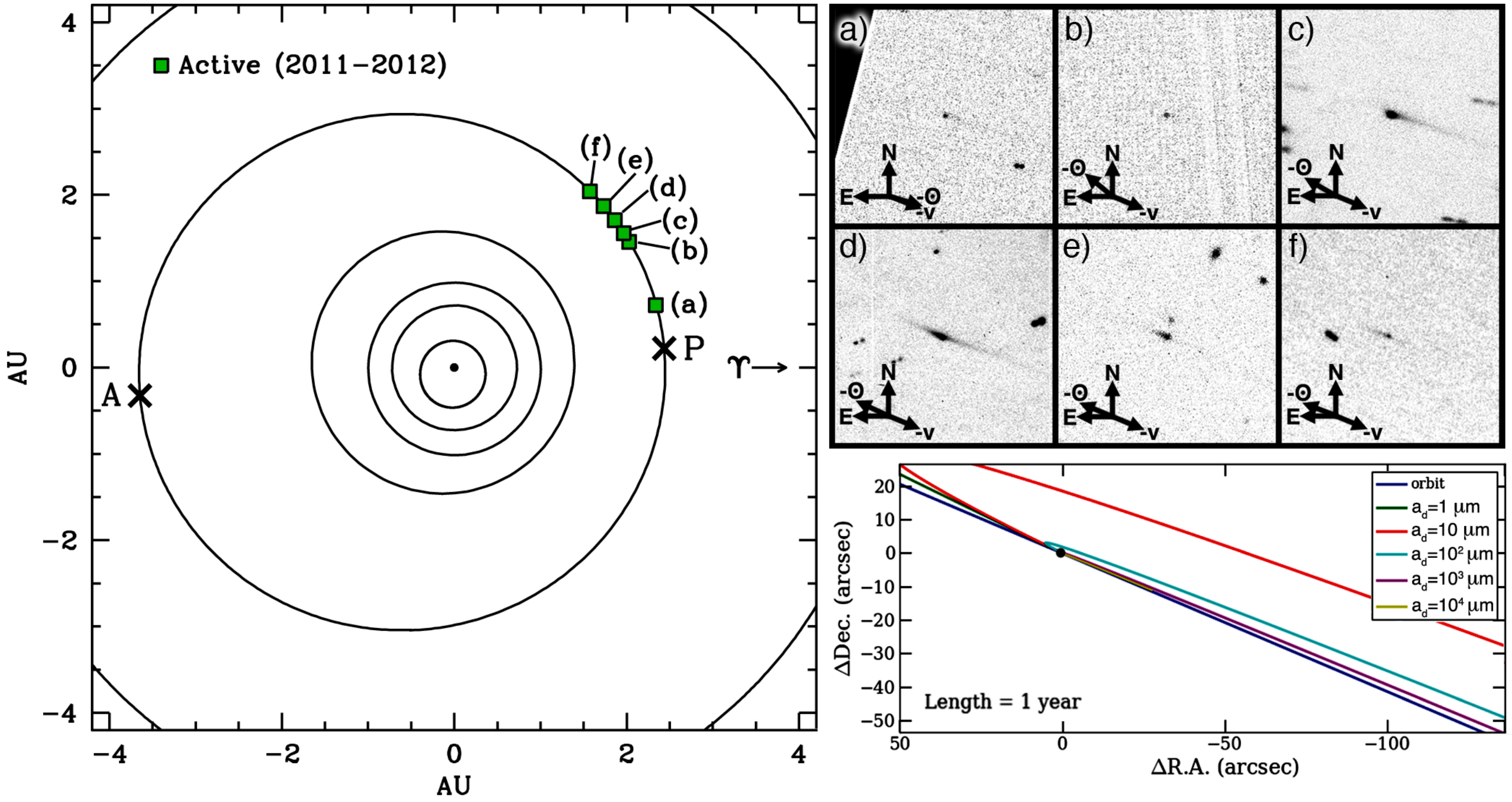}
\caption{\small 
Orbital position plot (left) and composite images (upper right) constructed from observations detailed in Table~\ref{obslog}, with the Sun (black dot) at the center, and the orbits of Mercury, Venus, Earth, Mars, P/2006 VW$_{139}$, and Jupiter shown as black lines.  Perihelion (P) and aphelion (A) are marked with crosses.  Plotted positions and images correspond to observations from (a) 2011 August 30, (b) 2011 November 5, (c) 2011 November 12-14 (November 14 UH composite image shown), (d) 2011 November 22 - December 4 (December 4 NTT composite image shown), (e) 2011 December 16-19 (December 19 UH composite image shown), and (f) 2012 January 7.  Each panel is $90''\times90''$ with the object at the center, and North (N), East (E), the antisolar vector ($-\odot$), and the negative heliocentric velocity vector ($-v$) marked with arrows.  Also shown is a syndyne plot (lower right) for the 2011 November 22 - December 4 period for a range of dust grain radii.
  }
\label{fig_images_orbplot}
\end{figure}

The morphology observed for P/2006~VW$_{139}$ from mid-November 2011 through January 2012 (shortly after perihelion) remains approximately constant (Figure~\ref{fig_images_orbplot}).  We observe the presence of a short dust tail ($\sim10''$) pointed in the antisolar direction, a longer dust trail ($\sim60''$) aligned with the object's orbit plane, and a coma with an intrinsic FWHM of $\sim0\farcs7$ when deconvolved with the seeing.  This morphology is similar to that observed for MBC P/2010 R2 (La Sagra) \citep{hsi12b}, as well as other well-established comets \citep[e.g.\ 2P/Encke, C/Austin 1990 V;][]{lis98,rea00}.  Detailed dust modeling will be required to confirm the nature of the observed dust features, but a simple syndyne plot \citep[Figure~\ref{fig_images_orbplot}; cf.][]{fin68} indicates that the antisolar tail is likely composed of small particles with short dissipation times (requiring recent emission to still be present), while the orbit-aligned trail is composed of large particles with slow dissipation times (likely requiring months to traverse the observed distance from the nucleus).  The contemporaneous observation of both recent and old dust emission implies the action of a prolonged event, consistent with sublimation and inconsistent with an impulsive impact.  We furthermore compute that particles with radii of $a_d=0.1~{\rm \mu m}$ should move $10''$ from the nucleus in just 22 days, creating an observable gap between the antisolar tail and the nucleus.  No such gap is observed between November and January, implying continuous replenishment of the tail over this period, again consistent with cometary activity.

Disrupted asteroid (596) Scheila exhibited a 30\% decline in coma brightness in 8 days \citep{jew11}.  Meanwhile, other established MBCs exhibited steady or increasing coma brightness over longer periods of time \citep[cf.][]{hsi12b}.  Photometry of P/2006~VW$_{139}$ (Table~\ref{obslog}) shows no decline in near-nucleus brightness over 29 days between November and early December (though does show a 40\% decline over 34 days between December and January).  Coupled with our morphological analysis, the coma's photometric behavior over this period provides further support for the observed activity's sublimation-driven nature.

For reference, we compute the dust-to-nucleus scattering surface area ratio, $A_d/A_N$, determined for other MBCs \citep[cf.][]{hsi12b}.  Lacking independent constraints at this time, we use the absolute $R$-band magnitude, $H_R=16.4$~mag, computed by \citet{spa09} only using observations obtained well before activity was discovered, and find $A_d/A_N\sim2.5$ during the object's peak observed activity in November.  This level of dust production relative to nucleus size is comparable to that of 133P and 176P, but an order of magnitude lower than for other MBCs, 238P, P/2008 R1 (Garradd), and P/La Sagra \citep{hsi12b}.  Multi-filter NTT observations (Table~\ref{obslog}) indicate that the near-nucleus coma has an approximately Solar color of $B-R=1.06\pm0.04$, similar to other MBCs \citep{jew09,hsi09a,hsi10,hsi11a,hsi12b}

\subsection{Spectroscopic Analysis\label{spec_results}}

\begin{figure}
\includegraphics[width=4.0in]{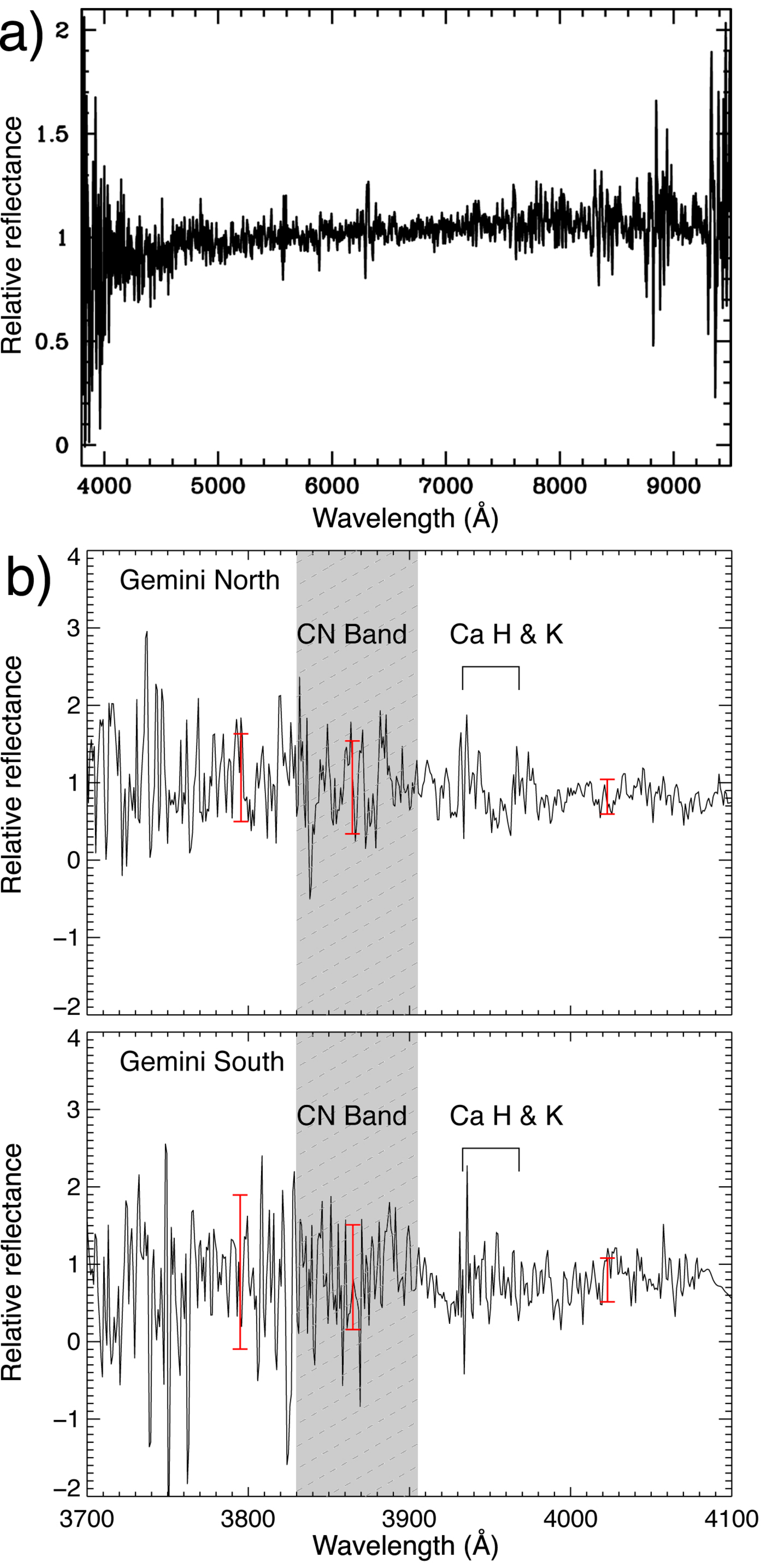}
\caption{\small 
(a) Relative reflectance spectrum of P/2006 VW$_{139}$ from the WHT.
(b) Relative reflectance spectra of P/2006 VW$_{139}$ from GN (upper panel) and GS (lower panel).  Shaded regions indicate the wavelength region where the CN emission band is expected.  Red error bars show 1-$\sigma$ uncertainties in the CN band region and in adjacent wavelength regions.
  }
\label{fig_spec}
\end{figure}

WHT spectroscopy was extracted using an aperture $10''$ long along the slit centered on the target. A spectrophotometric standard, Feige 15, and a G2V solar analog star, HD28099, were observed to allow removal of atmospheric absorption features and calculation of the relative reflectance spectrum.  The resulting spectrum (Figure~\ref{fig_spec}a) is approximately linear with a red slope of 7.2\%/1000\AA, similar to that seen in other cometary dust comae \citep{kol04}.
% Unfortunately, we did not obtain enough signal-to-noise along the length of the tail to permit spatially-resolved measurements.

Gemini spectroscopy was extracted using an aperture $5\farcs1$ long along the slit.  A G5V solar analog star, HD10097, was observed for approximate flux calibration and to calculate the relative reflectance spectra.  The resulting GN and GS spectra (Figure~\ref{fig_spec}b) are very similar.  Given its slightly higher signal-to-noise ratio, however, we use the GN spectrum to compute gas production rates.

We measure the standard errors in three wavelength regions $70~{\rm \AA}$ in width and adopt the largest $\sigma$ of 0.057 as the observational uncertainty of the violet CN emission band.
%The B-band magnitude of HD 10097 is 9.81 and the estimated flux density of 300163 in the Johnson B-band is 2.26 x 10$^{-16}$ erg cm$^{-2}$ s$^{-1}$ \AA$^{-1}$.
We then employ a simple \citet{has57} model to derive the CN production rate \citep[cf.][]{hsi12a,jew12a}, using a resonance fluorescence efficiency of $g[1{\rm AU}]=3.63\times10^{-13}$~erg~s$^{-1}$~molecule$^{-1}$ \citep{sch10}. 
We find an upper limit to the CN production rate of $Q_{\rm CN}<1.3\times10^{24}$~mol~s$^{-1}$.  No physical constraints on the CN to water production rate ratio in MBCs are currently available, but adopting the average ratio in other observed comets ($\log [{Q_{\rm CN}}/{Q_{\rm OH}}]=-2.5$; ${Q_{\rm OH}}/{Q_{\rm H_{2}O}} = 90$\%)
\citep[cf.][]{hsi12a}, we infer a water production rate of $Q_{\rm H_{2}O}<10^{26}$~mol~s$^{-1}$.

\subsection{Dynamical Analysis\label{dyn_results}}

To ascertain whether P/2006~VW$_{139}$ is likely to be native to its current location, we analyze its dynamical stability \citep[cf.][]{jew09,hsi12a,hsi12b}.  We produce nine randomly-generated sets of 100 Gaussian-distributed massless test particles in orbital element space, centered on the object's current osculating orbital elements, where three sets are characterized by $\sigma$ values of $\sigma_a=0.0001$~AU, $\sigma_e=0.0001$, and $\sigma_i=0.001\degr$.  These $\sigma$ values are chosen to adequately explore the stability of the region.  As a numbered asteroid, P/2006 VW$_{139}$'s true orbital element uncertainties are actually $\sim100-1000\times$ smaller.  Another three sets are characterized by $\sigma$ values $10\times$ as large, and the final three sets are characterized by $\sigma$ values $100\times$ as large.  Treating the eight major planets as massive particles, we then use the N-body integration package, Mercury \citep{cha99}, to integrate the orbits of these test particles forward in time for 100 Myr.  Limitations on computing resources unfortunately prevent us from conducting significantly longer integrations (e.g., $\geq1$~Gyr) in a reasonable amount of time.

\begin{figure}
\includegraphics[width=5.5in]{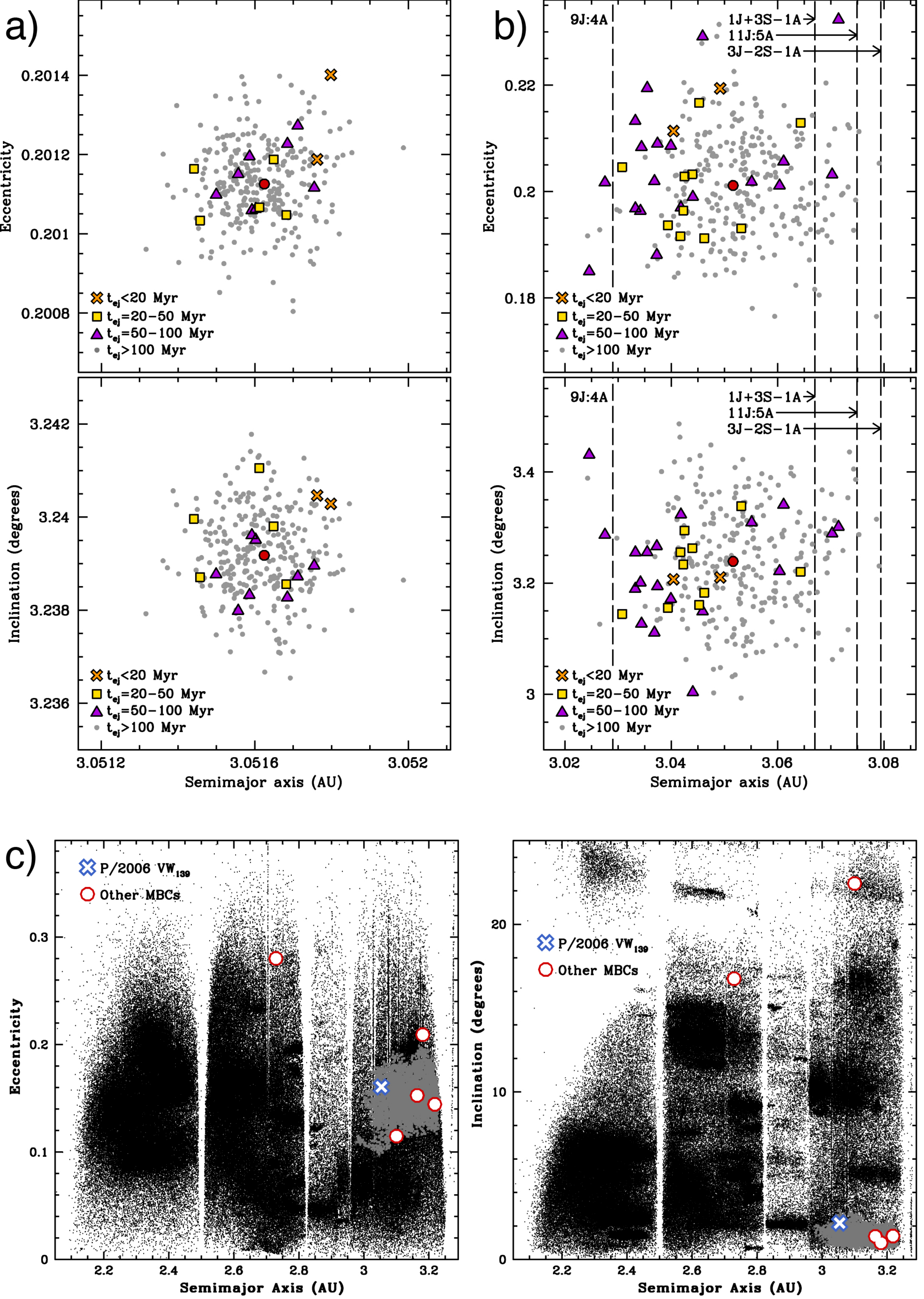}
\caption{\small 
(a) Plots of semimajor axis versus eccentricity (top) and inclination (bottom) showing initial osculating elements of three $1-\sigma$ sets of Gaussian-distributed test particles whose orbits are integrated for 100 Myr (Section~\ref{dyn_results}), with the current osculating orbital elements of P/2006~VW$_{139}$ shown as a red circle at the center of each panel. Particles ejected in $<$20 Myr, between 20 Myr and 50 Myr, and between 50 Myr and 100 Myr are plotted with orange X symbols, yellow squares, and purple triangles, respectively, while particles that are not ejected within 100 Myr are marked with grey dots.
(b) Same as (a) but for $100-\sigma$ test particle sets, where the positions of the 9J:4A, 1J+3S-1A, 11J:5A, and 3J-2S-1A mean motion resonances are marked with dotted lines.
(c) Plots of proper semimajor axis vs. proper eccentricity (left) and proper inclination (right) of main-belt asteroids (small black dots), Themis family members (small gray dots), P/2006~VW$_{139}$ (blue X symbol), and other MBCs (red circles).
  }
\label{fig_dyn}
\end{figure}

We find that 5\% of the 1-$\sigma$ test particles, 3\% of the 10-$\sigma$ test particles, and 10\% of the 100-$\sigma$ test particles exceed a heliocentric distance of $50$~AU, and are effectively ejected from the asteroid belt, within 100~Myrs (Figure~\ref{fig_dyn}).  P/2006 VW$_{139}$ itself is found to be stable over the 100~Myr test period (neglecting non-gravitational forces), consistent with its Lyapunov time of $T_{\rm lyap}>1$~Myr, where a body is considered stable if $T_{\rm lyap}>100$~kyr \citep{tsi03}.  No systematic distribution is evident for ejected 1-$\sigma$ or 10-$\sigma$ test particles, though 44\% of 100-$\sigma$ test particles with $a<3.04$~AU are ejected, perhaps due to the 9J:4A mean-motion resonance with Jupiter at 3.029~AU (Figure~\ref{fig_dyn}b).  For reference, P/2006~VW$_{139}$ fluctuates between $a=3.047$~AU and $a=3.069$~AU over the course of our simulations.

As two MBCs (133P and 176P) belong to the $\sim$2.5~Gyr-old \citep{nes03} Themis asteroid family, and 133P additionally belongs to the young $<10$~Myr Beagle sub-family \citep{nes08}, we perform a search for any family associations that P/2006~VW$_{139}$ may have.  Employing Hierarchical Clustering Method \citep[HCM;][]{zap90} analysis and analytically-determined proper orbital elements retrieved on 2011 December 1 from AstDyS ({\tt http://hamilton.dm.unipi.it/astdys/}), we compute the number of asteroids dynamically linked to P/2006~VW$_{139}$ as function of cut-off distance, $d_c$ (in velocity space).  At $d_c=63~{\rm m~s}^{-1}$, we find a statistically significant clustering of 24 asteroids, which then merges with the Themis family at $d_c=70~{\rm m~s}^{-1}$ (Figure~\ref{fig_dyn}c).  This possible P/2006 VW$_{139}$ sub-family is separated from the main Themis family by several two- and three-body mean-motion resonances (11J:5A, 3J-2S-1A, and 1J+3S-1A) with Jupiter and Saturn, though our dynamical simulations indicate these resonances may be only weakly destabilizing (Figure~\ref{fig_dyn}b).  P/2006 VW$_{139}$'s association with the Themis family is interesting because of the family's aforementioned association with 133P, 176P, and possibly 238P \citep{hag09}, though more detailed study is needed to clarify the nature of this association, as well as the significance of the possible sub-family.

\section{DISCUSSION\label{discussion}}

While spectroscopy did not reveal any evidence of gas emission (Section~\ref{spec_results}), morphological and photometric analyses (Section~\ref{morph_phot_results}) strongly suggest that the activity of P/2006~VW$_{139}$ is cometary in nature, and not impact-driven.  We conclude that this object is likely a MBC, making it the sixth such object discovered to date.  The discovery circumstances of the known MBCs suggest that many more must exist, particularly in the outer main belt \citep{hsi09}.  Therefore, we expect that current and next-generation all-sky surveys like Pan-STARRS will reveal more of the true extent of the population in the coming years, provided that techniques can be deployed to efficiently screen for such objects amid the enormous amounts of data produced by these surveys.  Thorough physical and dynamical investigations of each new discovered object will then be essential for first determining the most likely cause of the observed activity \citep[cf.][]{hsi12a,jew12b}, and then ascertaining the global properties of the population of objects confirmed as MBCs to better understand their implications for understanding terrestrial water delivery.

Additional observations of P/2006 VW$_{139}$ itself are encouraged until the end (in March 2012) of the 2011-2012 observing window to monitor the decline of activity for comparison to other MBCs \citep[cf.][]{hsi11b,hsi12b}.  Observations during the 2012-2013 observing window, when the object may be largely inactive, should be useful for determining its nucleus properties.  By adding to our knowledge of the properties of inactive MBC nuclei \citep[cf.][]{hsi09b,lic11}, the latter observations will help us better understand the relationship of active MBCs to the inactive main-belt population, perhaps facilitating the development of more powerful search methods, such as targeted monitoring of extremely likely MBC candidates.  Furthermore, robust protocols for identifying icy objects, even in the absence of activity (which is transient even for known MBCs), will be crucial for ascertaining the true abundance and distribution of ice in the asteroid belt.

%Comparison of P/2006 VW$_{139}$'s nucleus properties to those of other MBC nuclei as well as those of other asteroids \citep[e.g.,][]{hsi09,lic11} should then provide interesting insights

\begin{acknowledgements}
We appreciate comments from an anonymous referee that improved this manuscript.  H.H.H.\ is supported by NASA through Hubble Fellowship grant HF-51274.01 awarded by the Space Telescope Science Institute, which is operated by the Association of Universities for Research in Astronomy (AURA) for NASA, under contract NAS 5-26555.
B.Y., N.H., H.M.K., and K.J.M. acknowledge support through the NASA Astrobiology Institute under Cooperative Agreement NNA08DA77A.
B.N.\ is supported by the Ministry of Education and Science of Serbia (Project 176011).
A.F.\ is supported by the Science \& Technology Facilities Council (Grant ST/F002270/1).
M.S.K.\ is supported by NASA Planetary Astronomy Grant NNX09AF10G.
PS1 is operated by the PS1 Science Consortium and its member institutions, and also funded by NASA Grant NNX08AR22G issued through the NASA Science Mission Directorate's Planetary Science Division.
Gemini is operated by AURA under a cooperative agreement with the National Science Foundation (NSF) on behalf of the Gemini partnership.
The WHT is operated by the Isaac Newton Group in the Observatorio del Roque de los Muchachos of the Instituto de Astrof\'isica de Canarias.
The Faulkes Telescopes are operated by Las Cumbres Observatory Global Telescope Network.
SDSS-III ({\tt http://www.sdss3.org/}) is funded by the Alfred P. Sloan Foundation, the Participating Institutions, NSF, and the U.S.\ Department of Energy Office of Science, and managed by the Astrophysical Research Consortium for the SDSS-III Collaboration.
%The PS1 Surveys have been made possible through contributions of the Institute for Astronomy, the University of Hawaii, the Pan-STARRS Project Office, the Max-Planck Society and its participating institutes, the Max Planck Institute for Astronomy, Heidelberg and the Max Planck Institute for Extraterrestrial Physics, Garching, The Johns Hopkins University, Durham University, the University of Edinburgh, Queen's University Belfast, the Harvard-Smithsonian Center for Astrophysics, and the Las Cumbres Observatory Global Telescope Network, Incorporated, the National Central University of Taiwan, and the National Aeronautics and Space Administration under Grant No. NNX08AR22G issued through the Planetary Science Division of the NASA Science Mission Directorate.
\end{acknowledgements}

\begin{deluxetable}{lcrrcrrrrrrrcc}
%\footnotesize
\scriptsize
\tablewidth{0pt}
\tablecaption{Observations\label{obslog}}
\tablecolumns{14}
\tablehead{
\colhead{UT Date}
 & \colhead{Tel.\tablenotemark{a}}
 & \colhead{N\tablenotemark{b}}
 & \colhead{t\tablenotemark{c}}
 & \colhead{Filter}
 & \colhead{$R$\tablenotemark{d}}
 & \colhead{$\Delta$\tablenotemark{e}}
 & \colhead{$\alpha$\tablenotemark{f}}
 & \colhead{$\nu$\tablenotemark{g}}
 & \colhead{PA$_{-\odot}$\tablenotemark{h}}
 & \colhead{PA$_{-v}$\tablenotemark{i}}
 & \colhead{$\alpha_{pl}$\tablenotemark{j}}
 & \colhead{$m(R,\Delta,\alpha)$\tablenotemark{k}}
 & \colhead{$m_R(1,1,0)$\tablenotemark{l}}
}
\startdata
2011 Jul 18 & \multicolumn{4}{l}{\it Perihelion.......................} & 2.438 & 2.293 & 24.6 &   0.0 & 249.3 & 247.3 &  $0.8$ & --- & --- \\
2011 Aug 30 & PS1   &  2 &    60 & $z_{P1}$ & 2.447 & 1.806 & 21.4 & 12.2 & 253.7 & 248.5 &  $1.7$ & 20.16$\pm$0.14 & 15.6$\pm$0.3 \\
2011 Nov 05 & PS1   &  2 &    90 & $i_{P1}$ & 2.496 & 1.517 &  4.6 & 30.7 &  50.5 & 247.9 &  $1.4$ & 19.00$\pm$0.09 & 15.1$\pm$0.2 \\
2011 Nov 12 & FTN   & 11 &   660 & $r'$     & 2.505 & 1.555 &  8.0 & 32.9 &  59.8 & 247.8 &  $1.1$ & 18.69$\pm$0.09 & 15.1$\pm$0.2 \\
2011 Nov 14 & UH    &  8 &  2400 & $R$      & 2.506 & 1.561 &  8.4 & 33.2 &  60.4 & 247.7 &  $1.1$ & 18.62$\pm$0.05 & 15.1$\pm$0.2 \\
2011 Nov 14 & FTN   &  3 &   540 & $R$      & 2.506 & 1.561 &  8.4 & 33.2 &  60.4 & 247.7 &  $1.1$ & 18.64$\pm$0.05 & 15.1$\pm$0.2 \\
2011 Nov 18 & PT    & 19 & 11100 & $R$      & 2.510 & 1.586 & 10.0 & 34.3 &  62.2 & 247.7 &  $0.9$ & 18.60$\pm$0.10 & 15.0$\pm$0.2 \\
%2011 Nov 19 & PT    &  8 &  4800 & $R$      & 2.512 & 1.592 & 10.5 & 34.5 &  62.6 & 247.6 &  $0.9$ & 18.60$\pm$0.10 & XXX \\
2011 Nov 19 & HCT   & 22 & 11400 & $R$      & 2.512 & 1.596 & 10.6 & 34.6 &  62.8 & 247.6 &  $0.9$ & 18.64$\pm$0.02 & 15.0$\pm$0.2 \\
2011 Nov 22 & WHT   &  3 &   780 & $r'$     & 2.516 & 1.621 & 11.9 & 35.5 &  63.8 & 247.6 &  $0.8$ & 18.89$\pm$0.02 & 15.1$\pm$0.2 \\
2011 Nov 22 & WHT   &  3 &  2700 & spec.    & 2.516 & 1.621 & 11.9 & 35.5 &  63.8 & 247.6 &  $0.8$ &  ---           &  --- \\
2011 Nov 30 & UH    &  5 &  3000 & $R$      & 2.525 & 1.685 & 14.4 & 37.4 &  65.4 & 247.5 &  $0.5$ & 19.04$\pm$0.02 & 15.1$\pm$0.2 \\
2011 Dec 02 & GN    &  8 &  4800 & spec.    & 2.527 & 1.704 & 15.0 & 38.0 &  65.7 & 247.5 &  $0.4$ &  ---           &  --- \\
2011 Dec 04 & NTT   &  2 &  1200 & $R$      & 2.530 & 1.724 & 15.6 & 38.5 &  66.0 & 247.4 &  $0.4$ & 19.12$\pm$0.03 & 15.1$\pm$0.2 \\
2011 Dec 04 & NTT   &  1 &   600 & $B$      & 2.530 & 1.724 & 15.6 & 38.5 &  66.0 & 247.4 &  $0.4$ & 20.18$\pm$0.03 &  --- \\
2011 Dec 05 & GS    &  6 &  3600 & spec.    & 2.531 & 1.735 & 15.9 & 38.8 &  66.2 & 247.4 &  $0.3$ &  ---           &  --- \\
2011 Dec 16 & FTS   & 12 &   720 & $r'$     & 2.546 & 1.861 & 18.7 & 41.7 &  67.4 & 247.4 &  $0.0$ & 19.70$\pm$0.09 & 15.3$\pm$0.2 \\
2011 Dec 19 & UH    &  3 &   900 & $R$      & 2.549 & 1.895 & 19.2 & 42.4 &  67.7 & 247.4 & $-0.1$ & 19.68$\pm$0.03 & 15.3$\pm$0.2 \\
2012 Jan 07 & LOT   &  8 &  1440 & $R$      & 2.577 & 2.152 & 21.7 & 47.4 &  69.2 & 247.6 & $-0.6$ & 20.43$\pm$0.10 & 15.7$\pm$0.2 \\
2014 Mar 13 & \multicolumn{4}{l}{\it Aphelion.........................} & 3.660 & 2.697 &  4.5 & 180.0 & 278.6 & 294.1 & $-1.2$ & --- & --- \\
2016 Nov 08 & \multicolumn{4}{l}{\it Perihelion.......................} & 2.436 & 1.823 & 21.3 &   0.0 &  65.1 & 248.2 &  $1.1$ & --- & ---
\enddata
\tablenotetext{a}{Telescope.}
\tablenotetext{b}{Number of exposures.}
\tablenotetext{c}{Total integration time, in s.}
\tablenotetext{d}{Heliocentric distance, in AU.}
\tablenotetext{e}{Geocentric distance, in AU.}
\tablenotetext{f}{Solar phase angle (Sun-object-Earth), in degrees.}
\tablenotetext{g}{True anomaly, in degrees.}
\tablenotetext{h}{Position angle of the antisolar vector, in degrees East of North.}
\tablenotetext{i}{Position angle of the negative velocity vector, in degrees East of North.}
\tablenotetext{j}{Orbit plane angle, in degrees.}
\tablenotetext{k}{Mean apparent magnitude in specified filter.}
\tablenotetext{l}{Absolute $R$-band magnitude (at $R=\Delta=1$~AU and $\alpha=0\degr$), assuming solar colors and IAU $H,G$ phase-darkening where $G=0.15$, where the listed uncertainty is dominated by the estimated uncertainty in $G$.}
\end{deluxetable}

\clearpage

\clearpage

\clearpage

\clearpage

\end{document}